\documentclass[amsmath,amssymb,twocolumn]{aastex631}
\usepackage{tabularx}
\usepackage{graphicx}
\usepackage{dcolumn}
\usepackage{bm}

\begin{document}

\title{Gravitational Waves Detected by a Burst Search \\ in LIGO/Virgo's Third Observing Run}

\author{T. Mishra}
\affiliation{Department of Physics, University of Florida, PO Box 118440, Gainesville, FL 32611-8440, USA}
\author{S. Bhaumik}
\affiliation{Department of Physics, University of Florida, PO Box 118440, Gainesville, FL 32611-8440, USA}
\author{V. Gayathri}
\affiliation{Leonard E. Parker Center for Gravitation, Cosmology, and Astrophysics, University of Wisconsin–Milwaukee, Milwaukee, WI 53201, USA}
\author{Marek J. Szczepa\'nczyk}
\affiliation{Faculty of Physics, University of Warsaw, Ludwika Pasteura 5, 02-093 Warsaw, Poland}
\author{I. Bartos}
\affiliation{Department of Physics, University of Florida, PO Box 118440, Gainesville, FL 32611-8440, USA}
\author{S. Klimenko}
\affiliation{Department of Physics, University of Florida, PO Box 118440, Gainesville, FL 32611-8440, USA}

\begin{abstract}
Burst searches identify gravitational-wave (GW) signals in the detector data without use of a specific signal model, unlike the matched-filter searches that correlate data with simulated signal waveforms (templates). While matched filters are optimal for detection of known signals in the Gaussian noise, the burst searches can be more efficient in finding unusual events not covered by templates or those affected by non-Gaussian noise artifacts. Here, we report the detection of 3 gravitational wave signals that are uncovered by a burst search Coherent WaveBurst (cWB) optimized for the detection of binary black hole (BBH) mergers. They were found in the data from the LIGO/Virgo's third observing run (O3) with a combined significance of 3.6\,$\sigma$. Each event appears to be a BBH merger not previously reported by the LIGO/Virgo's matched-filter searches. The most significant event has a reconstructed primary component in the upper mass gap ($m_1 = 70^{+36}_{-18}\,$M$_\odot$), and unusually low mass ratio ($m_2/m_1\sim0.3$), implying a dynamical or AGN origin. The 3 new events are consistent with the expected number of cWB-only detections in the O3 run, and belong to the stellar-mass binary population with the total masses in the $70-100$\,M$_\odot$ range.

\end{abstract}


\section{Introduction}

The LIGO \citep{TheLIGOScientific:2014jea} and Virgo \citep{TheVirgo:2014hva} detectors have so far reported the discovery of about 90 gravitational-wave sources \citep{LIGOScientific:2021psn}. Each detection has been a compact binary coalescence (CBC) of black holes and neutron stars. These discoveries already delivered a wealth of information about the formation and evolution of compact objects, including the existence of heavy ($\gtrsim 50$\,M$_\odot$) black holes that are difficult to explain with a stellar origin \citep{PhysRevLett.125.101102}, objects in the so-called lower-mass gap where Galactic observations suggested a dearth of black holes \citep{2020ApJ...896L..44A,2024ApJ...970L..34A}, and hints of orbital eccentricity  \citep{2022NatAs...6..344G,2020ApJ...903L...5R,2023NatAs...7...11G}.

Binary black hole (BBH) mergers can have a variety of origins that leave imprints on the binaries' properties. Possible origins include isolated formation from stellar binaries \citep{2012ApJ...759...52D, Belczynski:2016obo, deMink:2016vkw}, binary evolution of Pop III stars \citep{Tanikawa:2020abs, Tanikawa:2024mpj, Mestichelli2024}, dynamical formation in dense environments such as galactic nuclei or globular clusters \citep{OLeary:2005vqo, 2010MNRAS.402..371B, Rodriguez:2016kxx}, or gas capture in Active Galactic Nuclei \citep{2017ApJ...835..165B, Tagawa:2019osr}. Generally, isolated binaries are expected to be lighter, close to equal mass black holes with low spins aligned with the orbital angular momentum. While dynamical formation and gas-capture can produce heavier black holes through mass segregation and multiple consecutive mergers \citep{2017ApJ...840L..24F}, more asymmetric masses due to the quasi-random pairing of the binary partners, and possibly high spins misaligned from the binary orbit \citep{Rodriguez:2016vmx, Talbot:2017yur}. Dynamical formation and AGN gas-capture can also introduce orbital eccentricity as these binaries can interact soon before merger \citep{2022Natur.603..237S}.

Our ability to discover and identify black hole mergers depends mainly on the sensitivity of the GW detectors, but also on the utilized search algorithms. Most binary searches use a gravitational-wave template bank that densely covers the targeted binary parameter space with waveforms, and use matched filters to compare each of the waveforms to the recorded data to find the best match \citep{Usman:2015kfa, 2017PhRvD..95d2001M, Aubin:2020goo}. Such {\it template-based} searches are optimal in the presence of stochastic noise, and if the covered parameter space includes that of the binary merger in question. Template-based gravitational wave searches have been exceedingly successful, and in fact, all gravitational wave signals discovered to date have been identified or co-identified by the template-based searches.

The assumptions under which template-based searches are optimal do not always hold. First, detector noise includes non-Gaussian noise artifacts, a.k.a. glitches, that must be separated from real signals. This problem is particularly acute for the heaviest detectable black hole mergers whose waveforms are shorter and more similar to some of the glitches. Second, at present, template banks used in searches do not cover the full BBH parameter space. Orbital eccentricity is not included, and only black hole spins that align with the orbital axis are allowed \citep{2024PhRvD.109d4066S}. For binaries with eccentric orbits or highly precessing black hole spins, this can substantially reduce the search sensitivity \citep{2024PhRvD.110b3038S,PhysRevD.102.043005}. Also, templates are lacking the sub-dominant modes that may compromise signal detection~\citep{PhysRevD.89.102003,PhysRevD.90.124004,PhysRevD.93.084019,PhysRevD.95.104038}.

To overcome the challenges of incomplete parameter space coverage and non-Gaussian detector noise, burst searches have been developed that do not closely rely on precise gravitational waveforms, but instead use more general assumptions about the signal and the background, e.g., by looking at excess energy coherently deposited in multiple gravitational wave detectors. A prominent model-agnostic search algorithm is {\it coherent WaveBurst} (cWB; \cite{Klimenko:2008,Klimenko:2016}). While cWB can detect binary mergers across the full parameter space accessible by the LIGO and Virgo detectors, its relative sensitivity compared to template-based methods makes it the most relevant for the heaviest black holes ($\gtrsim20$\,M$_\odot$; \cite{2022A&A...659A..84A}), and in the case of unusual binary black holes not covered by template-based searches, such as eccentric and precessing mergers \citep{LIGOScientific:2023lpe, 2024PhRvD.110d4013G}.

cWB has been in use searching for binary mergers in all data recorded by LIGO/Virgo so far, with remarkable results. Among others, it was the first algorithm to detect gravitational waves \citep{2016PhRvL.116f1102A}, and identified the heaviest black hole merger to date, GW190521, with a much lower false alarm rate than template-based searches \citep{2020PhRvL.125j1102A}. cWB detected almost all heavy black hole mergers that were also reported by template-based searches \citep{2019PhRvX...9c1040A,2021PhRvX..11b1053A,2023PhRvX..13d1039A}. 

Nonetheless, so far, there has been no GW event that was only identified by cWB and not by any LIGO/Virgo's matched-filter search. In this paper, we report the results of a {\it revisited} search for binary black hole mergers using cWB during the third observing run (O3) of LIGO and Virgo. For this work, cWB has undergone substantial improvements compared to the previously used versions. With these changes, the more sensitive cWB search identified three new event candidates that are presented in the paper.

\section{Method}

cWB is an analysis algorithm used in searches for transient GW signals with the networks of GW detectors. Designed to operate without a specific waveform model, cWB identifies coincident excess power in the multi-resolution time-frequency (TF) representations of the detector strain data \citep{Klimenko:2008, Klimenko:2016}, for signal frequencies up to a few kHz and duration up to a few seconds. After the excess power events are identified in the TF data, the cWB reconstructs the source sky location and signal waveforms recorded by the detectors with the constrained maximum likelihood method \citep{PhysRevD.72.122002,Klimenko:2016}. The cWB detection statistic is based on the coherent energy $E_c$ obtained by cross-correlating the signal waveforms reconstructed in the detectors and normalized by the spectral amplitude of the noise. The $\sqrt{E_c}$ is the lower bound on the event network SNR $\rho_N$ and used for the initial selection of events produced by cWB. The production rate of cWB events is dominated by the non-stationary detector noise (glitches) and further reduced with the post-production vetoes. One of the primary signal-independent veto statistic is the network correlation coefficient $c_c = E_c/(E_c+E_n)$, where $E_n$ is the normalized residual noise energy estimated after the reconstructed signal is subtracted from the data. Typically, for a GW signal $c_c \sim 1$, and for glitches $c_c < 1$. Therefore, the candidate events with low value of $c_c$ are rejected as potential glitches. For GW signals the residual energy $E_n$ is expected to follow the $\chi^2$ distribution with the number of degrees of freedom $N_{DoF}$ which is proportional to the number of the TF data samples comprising the event. The reduced $\tilde{\chi}^2=E_n/N_{DoF}$ statistic is used as a powerful veto to identify and remove glitches with $\tilde{\chi}^2$ significantly larger than 1.  
While cWB uses the model-agnostic detection, reconstruction, and veto algorithms, it can also employ a weak signal-dependent veto to improve the identification of BBH signals. It is based on a generic property of a BBH signal which merger frequency is $f_c\propto{1/M}$ and the slope of a chirp track in the TF domain is approximated as $\dot{f}\propto{{\cal M}^{5/3}_{\rm c}f^{11/3}}$~\citep{Tiwari_2016}, where $M$ and ${\cal M}_{\rm c}$ are the red-shifted total and chirp masses respectively. Both parameters can be estimated from the frequency evolution $f(t)$ of detected events and used to improve the separation of BBH events and glitches. 

All cWB events are ranked by the reduced coherent network SNR $\rho_\mathrm{cWB}$:
\begin{eqnarray}
\label{eq1}
\rho_\mathrm{cWB}=\sqrt{\frac{E_c}{1+\tilde{\chi}^2(max\{1,\tilde{\chi}^2\}-1)}} \; .
\end{eqnarray}
For estimation of the statistical significance of the GW candidates, they are ranked against background events triggered by the detector noise. The background data sample, equivalent to $O(1000)$ years of the observation time, is obtained by repeating the cWB analysis on the time-shifted data. To exclude astrophysical events from the background data, the time shifts are selected to be much larger than the expected signal time delay between the detectors. The statistical significance of each cWB event is quantified by the false-alarm rate (FAR) given by the rate of background triggers with values of $\rho_\mathrm{cWB}$ larger than the given event. To track possible variations of the detector noise the event significance is estimated on the nearby data intervals one-two weeks long.

\subsection{cWB upgrade}

There are three significant changes in the cWB algorithm with respect to the version used during the O3 observing run \citep{2023PhRvX..13d1039A,Drago:2020kic}. 

\begin{figure*}[t!]
    \includegraphics[width=1\textwidth]{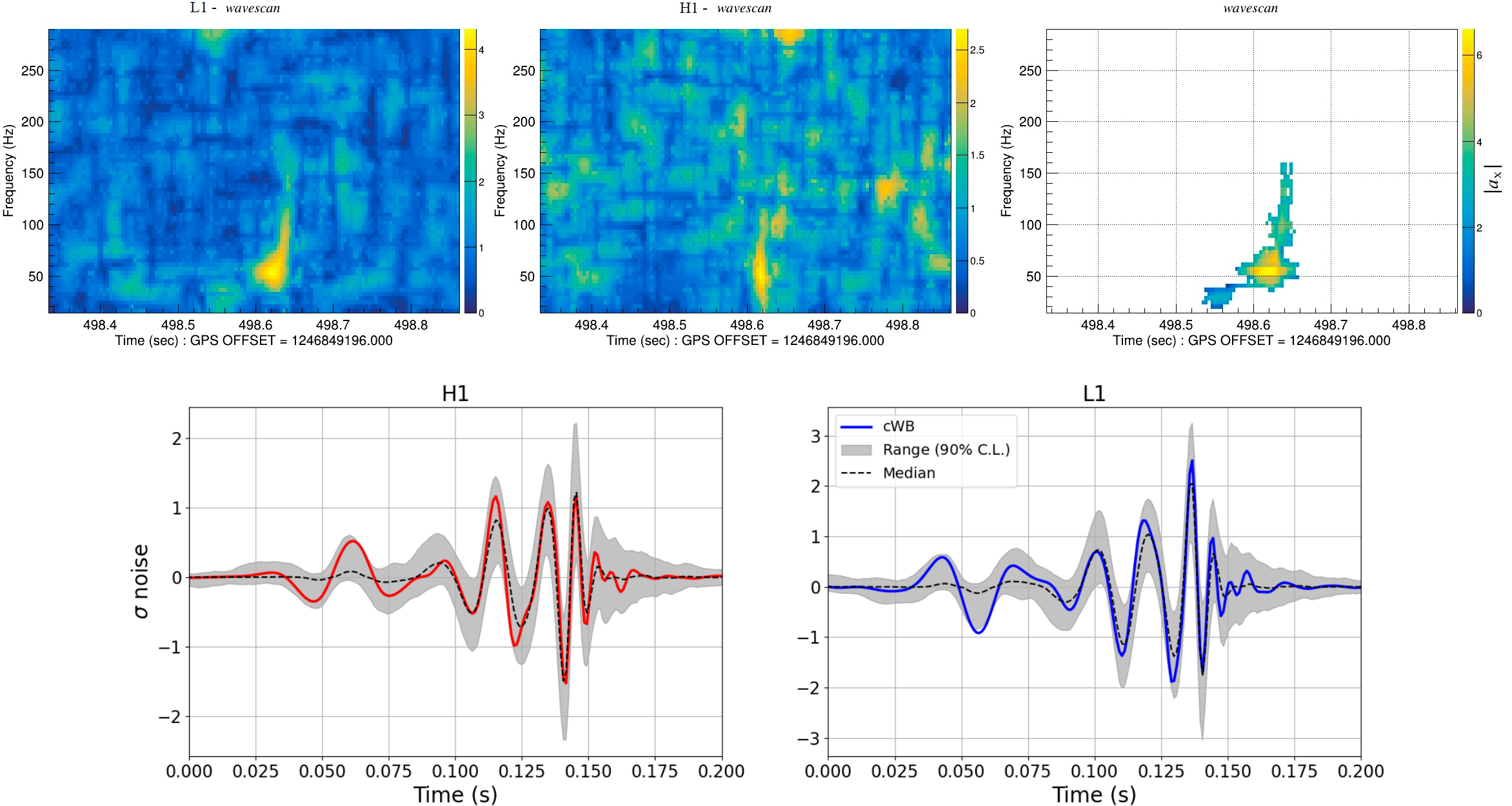}
    \caption{WaveScans for L1 (top left), H1 (top middle), and the cross-power TF map (top right) comprising the most significant event candidate 190711, followed by its reconstructed gravitational waveforms (bottom). The red and blue lines show the cWB waveform reconstruction for the H1 and L1 detectors respectively, together with their 90$\%$ confidence intervals (shaded regions calculated over the weighted parameter estimation samples injected into the off-source data and reconstructed with cWB).}
    \label{fig:event1}
\end{figure*}

First, the WDM wavelet transform \citep{Necula:2012zz} is replaced with the multi-resolution {\it WaveScan} transform based on the Gabor wavelets \citep{Klimenko:2022nji}, reducing the temporal and spectral leakage in the time-frequency data. 
For example, the high-resolution TF distribution provided by {\it WaveScan} in Fig.~\ref{fig:event1} shows the TF evolution of one of the GW events recorded by the LIGO-Livingston (L1) detector (left) and the LIGO-Hanford (H1) detector (center) on July 11, 2019 (event ID 190711).

Second, in addition to the excess-power statistic, the cross-power statistic \citep{Klimenko:2022nji} is used for the identification of transient events. The cross-power amplitude $a_\times$ of each TF data sample (or {\it WaveScan} pixel) is maximized over all possible time-of-flight delays of a GW signal in the detector network. Both statistics follow a predictable half-normal distribution with the unity variance expected for the quasi-stationary detector noise (see \cite{Klimenko:2022nji}). All TF pixels with the excess-power amplitude $a_e>a_o$ are selected for the analysis, where in this search $a_o=2.3$. Operation at the lower value of $a_o$ is limited by the computing requirements and by the execution time. The nearby TF pixels are clustered to form the initial cWB events. Fig.~\ref{fig:event1} (right) shows the cross-power distribution for the event 190711 identified by cWB. The clustered excess-power $p_e=\sum_{i=1}^m{a^2_e[i]}$ and the cross-power $p_\times=\sum_{i=1}^m{a^2_\times[i]}$ provide an estimator of the event SNR $\rho_e$ and the coherent SNR $\rho_\times$ respectively
\begin{eqnarray}
\label{eq2}
&\rho_e=2\sqrt{2f_{os}}(\sqrt{p_e}-a_o\sqrt{m}) , \;  \\ 
\label{eq3}
&\rho_\times=\sqrt{f_{os}p_\times} \; ,
\end{eqnarray}
where $m$ is the number of the TF pixels comprising the event, and $f_{os}=1/32$ is the {\it WaveScan} oversampling factor used in this analysis. 
The $\rho_e$ is optimized for the identification of events due to fluctuations of the quasi-stationary detector noise dominating the initial cWB rate of $O(100~Hz)$. By using the high-order moments of the excess and cross power statistics: $u^2=\sum_{i=1}^m{a^2_\times[i]a^2_e[i]}$ and $v=\sum_{i=1}^m{a_\times[i]a_e[i]}$, the $\rho_e$ is corrected by a factor 
\begin{eqnarray}
\label{eq4}
\delta=\frac{4}{\sqrt{mf_{os}}}\frac{r^2_\times}{r^2_s} \; , r^2_\times = \frac{p_\times}{p_e} \;, r_s = \frac{v^2+mu^2}{2p_e{p_\times}}. 
\end{eqnarray}
All events with $\rho_e-\delta<4$ are removed from the analysis.
The remaining cWB events, excluding GW signals, are mostly produced by the non-stationary detector noise with a typical rate of $O(0.1~Hz)$. After selection, the initial TF clusters are aggregated if they are within the time and frequency intervals of $0.25~s$ and $64~Hz$ respectively. It improves the collection of energy for transient events that can be fragmented into clusters. The glitch rate is reduced by applying a threshold on the upper bound of the coherent network SNR calculated for the defragmented events:
\begin{eqnarray}
\label{eq5}
\max\{\rho_\times,(\rho_e-\delta)\cdot\min(1,r^2_\times)\}\cdot{r_s} > 7.
\end{eqnarray}
The above conditions on the $\rho_e$ and $\rho_\times$ require that the GW signal fragments with SNR$>4$ and GW events with SNR$>7$ are accepted for the analysis. The reduced event rate ($O(0.01~Hz)$) makes the likelihood analysis and reconstruction of the remaining events computationally feasible. At this stage, the sky location of each event is determined and the signal waveforms are reconstructed with the inverse {\it WaveScan} transform \citep{Klimenko:2022nji}. The cWB events with  $\rho_\mathrm{cWB}>7$ are stored for the post-production analysis.

Finally, we substitute the post-production veto analysis with the machine-learning algorithm XGBoost. By using an ensemble of decision trees, it performs a classification of the event summary statistics produced by cWB ($E_c, c_c, \tilde{\chi}, \rho_e, \rho_\times, r_\times, r_s, f_c, {\cal M}_{\rm c}$, etc.) to improve separation of the genuine GW events and glitches. The decision trees are trained on a representative set of simulated BBH events (section \ref{subsec:sim}) and background events. Due to a weak dependence of the event summary statistics on the BBH dynamic, the algorithm retains a high detection efficiency for BBH signals outside of the training set.  A detailed description of the XGBoost post-production analysis is published elsewhere~\citep{PhysRevD.104.023014, PhysRevD.105.083018, PhysRevD.107.062002}. The final cWB detection statistic used for the estimation of the detection significance is the same as the one described in \cite{PhysRevD.105.083018}
\begin{equation}\label{eq6}
\rho_\mathrm{r} = \rho_\mathrm{cWB}\cdot W_{\mathrm{XGB}}, 
\end{equation}
where $W_{\mathrm{XGB}}$ is the XGBoost classification factor distributed between 0 (glitch) and 1 (signal) \citep{PhysRevD.104.023014}.

\subsection{Simulations}\label{subsec:sim}
The following simulation data sets are considered in this study: (i) stellar-mass BBH with aligned spins and quasi-circular orbits (the same as in \cite{PhysRevD.105.083018}), (ii) intermediate mass black hole (IMBH) mergers with the component masses in the range \{90 M$_\odot$ - 600 M$_\odot$\} and isotropic spin orientations (part of the GWTC-3 data release \cite{GWTC3_data_release}), (iii) BBH systems with the isotropic spin orientations and a wide range of masses following the population distribution described in \cite{2023PhRvX..13d1039A} (also part of the GWTC-3 data release \cite{GWTC3_data_release}). For all simulated signals, the redshift $z$ is drawn from a uniform distribution in the co-moving volume.

Simulated signals are injected into the detector data and recovered with the cWB pipeline. The simulation sets (i) and (ii), mixed in equal proportions, are used for the XGBoost training. The simulation set (iii) is used to estimate the probability of the astrophysical origin for detected events and the cWB sensitivity studies. From these studies, we observe a $\sim{39}\%$ improvement in the cWB detection efficiency as compared to the previous cWB search. For the rest of this paper, the simulation set (iii) has been re-weighted~\citep{Tiwari:2017ndi, PhysRevD.100.043030} to follow the POWER LAW+PEAK (PLP) astrophysical population model provided in GWTC-3 \citep{2023PhRvX..13a1048A}.

\section{Results}
In this section, we present the results from the reanalysis of the O3 data with the upgraded cWB pipeline. Unlike in the previous cWB run~\citep{2023PhRvX..13d1039A}, which employed two different search configurations, the updated cWB performed a single search for both stellar-mass BBH and IMBH mergers in the detector frequency band of $16-320$~Hz. Employing a single search reduces the data processing time and eliminates the need for a trial factor. Due to the variation of the detector noise in the first (O3a) and the second (O3b) halves of the O3 run, the two XGBoost configurations were trained separately for the O3a and O3b data. The training procedure was similar to the method described in \cite{PhysRevD.105.083018}, with the updated list of the input summary statistics produced by cWB.

\subsection{Detections}

In Table~\ref{tab:O3events} we report 33 GW events that are present in the GWTC-3 catalog \citep{2023PhRvX..13d1039A}, with an inverse false alarm rate (IFAR) of 1 year as the threshold for detection. In this search
we detect 9 GW events that were previously missed by the O3 cWB search. In addition to the known GW events, cWB identified 3 new event candidates that were not previously reported in the GWTC-3 catalog. The GPS time and IFAR of the cWB-only event candidates, as well as their signal-to-noise ratios (SNR) and the probability of astrophysical origin ($P_\mathrm{astro}$) are shown in Table \ref{tab:newevents}. The highest significance candidate has an IFAR $>40$ yr, placing it well above the detection threshold, while the lowest significance GW candidate, with IFAR $=4.5$ yr, is less clearly a detection. Nevertheless, the three new candidates together have a combined significance of 3.6\,$\sigma$, representing a statistical excess of events detected only by the burst search.

{\renewcommand{\arraystretch}{1.4}
\begin{table}
\begin{center}
\begin{tabular}{ c | c| c | c | c }
  \,\,Event ID\,\, & {\bf IFAR}  & \,\,SNR\,\, & \,\,$P_{\mathrm{astro}}$\,\, & \,GPS time \\
     &  [yr]  &  &  &  \\
\hline
    190711  &  {\bf 40.7} & 10.1 & 0.99 & 1246849694.6 \\
    190607  &  {\bf 11.7} & 9.4 & 0.99 & 1243931925.9 \\
    200318  &  {\bf 4.5} &  8.4 & 0.94 & 1268594035.1 \\
\end{tabular}
\end{center}
\caption{New GW events identified by the upgraded cWB search in the O3 data.}
\label{tab:newevents}
\end{table}
}

\subsection{Event properties}

To quantitatively assess the binary origin of the three cWB candidates, we carried out the binary parameter estimation  (PE) for each of the detected events using the Bilby framework \citep{2019ApJS..241...27A}.  Here, we compare model waveform templates with the observed data to compute the posterior probability for a specific set of parameters. The PE algorithm assumes that the detector noise is Gaussian and stationary. The analysis employed the IMRPhenomXPHM waveform model \citep{Pratten:2020ceb}, which describes the inspiral, merger, and ringdown phases of the BBH coalescence, incorporating precession and higher-order modes. For the parameter estimation analyses, we set the uniform chirp mass prior as \{ 10.8 - 90.8 \} $M_{\odot}$, uniform mass ratio prior as \{ 0.05 - 1 \}, uniform dimensionless spin magnitude priors as \{ 0 - 0.99 \} for both black holes in the binary, and sine priors for the tilt angles of the spins relative to the orbital angular momentum. Uniform priors ranging from 0 to $2\pi$ with periodic boundaries were used for the azimuthal angles between the spin vectors and the total angular momentum. We also set a uniform prior on the binary's luminosity distance in the source frame, with a range from 1\,Mpc to 10\,Gpc.

\begin{figure*}[hbt!]
    \centering
    \includegraphics[height=3in]{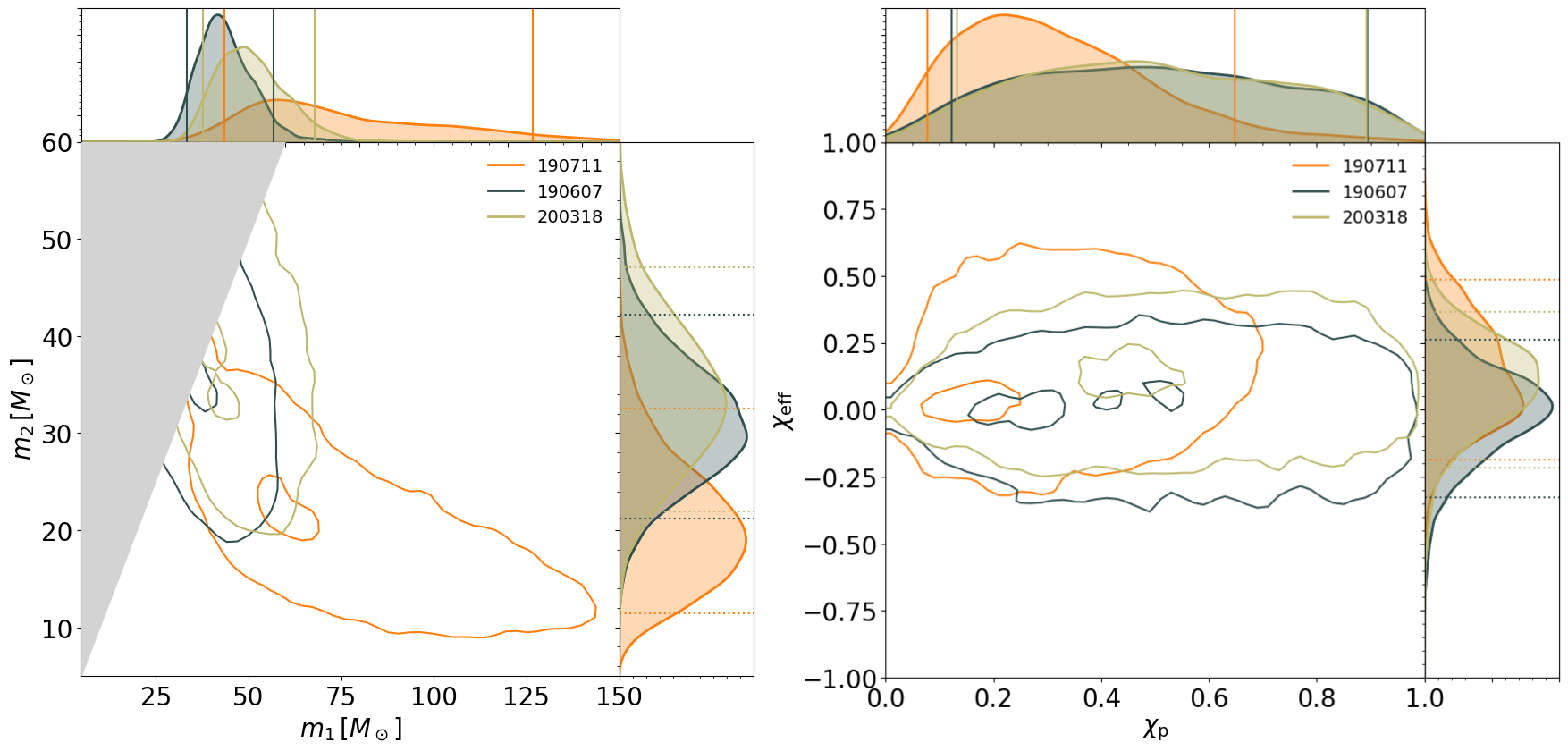}
    \caption{Posterior distributions: (Left) component mass distributions ($m_1$ and $m_2$) in the source frame, (Right) the effective spin ($\chi_{\mathrm{eff}}$) and effective precession spin distribution ($\chi_\mathrm{p}$) for the three new O3 events. The 90\% and 10\% credible regions are indicated by the solid contour in the joint distribution and by solid vertical and horizontal lines in the marginalized distributions.} 
    \label{fig:pe}
\end{figure*}

The parameter estimation results for the cWB candidates are reported in  Table \ref{tab:PE}. Our parameter estimation gives a good match for each of the candidates, as measured in the Bayes Factor (BF) of the best matching waveform for the events ($\ln(\mathrm{BF}) > 18$), indicating that a stellar-mass binary merger is an adequate explanation for each of the events. Fig \ref{fig:pe} (left), shows the 1D \& 2D posterior distribution for the component masses $m_1$ and $m_2$ of the source for each event with the IMRPhenomXPHM model. Fig \ref{fig:pe} (right) shows the 1D \& 2D posterior distribution for effective spin $\chi_{\mathrm{eff}}$ and effective precession spin $\chi_\mathrm{p}$ for each event with the IMRPhenomXPHM model. Based on these results, there is no evidence for precession. To confirm this finding, we also conducted parameter estimation using the aligned spin option of the IMRPhenomXPHM model. We observed that the BF ratio between the aligned and precessing spin results is approximately 1. 

To assess how well the signal model fits the data, we conducted a waveform reconstruction study using the PE waveform samples \citep{PhysRevD.100.042003,Szczepanczyk:2020osv,Gayathri:2020coq}. The PE samples were randomly selected from the posterior distribution for a given event. They were injected into the same data segment, within a time interval of 150 seconds. The data was subsequently analyzed with cWB. The reconstructed cWB waveforms were aligned in time to create the confidence intervals, demonstrating the expected variation in the modelled signal amplitudes as detected by cWB. Fig \ref{fig:event1} (bottom) presents the results of the waveform reconstruction study for the 190711 event. The light gray bands represent the 90\% confidence intervals and the black curves represent the median of the simulated waveform distribution. The red and blue curves, depicting the 190711 waveform reconstruction with cWB for the H1 and L1 detectors respectively, are well within the PE confidence intervals. We found no evidence for eccentricity from the waveform reconstruction study. The PE sample waveforms closely match the cWB point estimates for all three events demonstrating that the data is described well by both the spin-aligned and precessing signal models. The detected events are consistent with the stellar-mass BBH population. However, the most significant candidate, 190711, has an unusually low mass ratio of $q\equiv m_2/m_1={0.29}^{+0.17}_{-0.13}$. Among the heavy mergers $M_{\rm tot}>70\,$M$_\odot$ reported by LIGO/Virgo/KAGRA to date, only one other event, GW190929\_012149, has a similarly low mass ratio.

{\renewcommand{\arraystretch}{1.5}
\begin{table*}
\begin{center}
\begin{tabular}{ c | c | c | c | c | c | c | c | c  } 
  \,\,\,\,\,Event ID\,\,\,\,\, & \,\,\,\,${\cal{M}_{\rm c}}[M_{\odot}]$\,\,\,\,  & \,\,\,\,$M[M_{\odot}]$\,\,\,\, & \,\,\,\,$m_1[M_{\odot}]$\,\,\,\, & \,\,\,\,\,\,$m_2[M_{\odot}]$\,\,\,\,\,\, & \,\,\,\,\,\,\,\,\,\,\,\,\,$q$\,\,\,\,\,\,\,\,\,\,\,\,\, & \,\,\,\,\,\,\,\,\,\,$\chi_{\rm eff}$\,\,\,\,\,\,\,\,\,\, & \,\,\,\,\,\,\,\,\,\,\,\,\,$\chi_{\rm p}$\,\,\,\,\,\,\,\,\,\,\,\,\, & \,\,\,\,$D_{\rm L}[Gpc]$\,\,\,\, \\ 
  \hline
    190711 & ${31.3^{+8.5}_{-6.0}} $ & $91.0^{+51.9}_{-22.2} $  & $69.9^{+56.8}_{-26.3}$ & $20.1^{+12.5}_{-8.7}$ & ${0.29}^{+0.38}_{-0.19}$ & $0.12^{+ 0.37}_{-0.30}$ & $ {0.29}^{+0.36}_{-0.21}$ & $ {2.3}^{+01.9}_{-1.1}$ \\
    190607 &  ${31.4}^{+8.7}_{-6.0}$ & ${74.0}^{+19.9}_{-13.2}$  & $43.1^{+13.7}_{-9.6}$ & $30.9^{+11.2}_{-9.7}$ & $0.74^{+0.23}_ {-0.32}$ & ${0.0}^{+0.26}_{-0.32}$ & ${0.49}^{+0.40}_{-0.37}$ & ${4.1}^{+2.9}_{-2.2}$ \\
    200318 &  ${35.0}^{+10.6}_{-7.2}$ &  ${83.0}^{+25.0}_{-15.5}$  & $49.9^{+17.9}_{-12.2}$ & $33.5^{+13.5}_{-11.6}$ & $0.69^{+0.27}_{-0.31}$ & $0.10^{+0.26}_{-0.32}$ & $0.49^{+0.40}_{-0.36}$ & $ 5.1^{+3.5}_{-2.8}$ \\
  \hline
  \hline
   GW170729 &  ${35.4}^{+6.5}_{-4.8}$ & ${83.7}^{+13.0}_{-12.0}$  & $51.0^{+13.9}_{-12.4}$ & $31.9^{+9.3}_{-9.6}$ & ${0.63}^{+0.32}_{-0.26}$ & ${0.37}^{+0.21}_{-0.25}$ & ${0.42}^{+0.34}_{-0.29}$ & ${2.8}^{+1.4}_{-1.4}$ \\
   GW190521 & ${69.2}^{+17.0}_{-10.6}$  & ${150}^{+29}_{-17}$ & $85^{+21}_{-14}$ & $66^{+17}_{-18}$ & ${0.79}^{+0.19}_{-0.29}$ & ${0.08}^{+0.27}_{-0.36}$ & ${0.68}^{+0.27}_{-0.36}$ & ${5.3}^{+2.4}_{-2.6}$ \\
\end{tabular}
\end{center}
\caption{Reconstructed parameters of the three cWB event candidates. The reconstruction was carried out with Bilby using the IMRPhenomXPHM waveform model, which incorporates the inspiral, merger, and ringdown phases of binary black hole coalescence, including precession and higher-order modes, but does not include orbital eccentricity. Columns include the binary chirp mass [${\cal M}_{\rm c}=(m_1m_2)^{3/5}(m_1+m_2)^{-1/5}$], total mass ($M=m_1+m_2$), $m_1$, $m_2$, mass ratio ($q=m_2/m_1$), effective spin ($\chi_{\rm eff}$), precessing spin ($\chi_{\rm p}$), and luminosity distance ($D_{\rm L}$, [Gpc]) for the match between data and the best matching IMRPhenomXPHM waveform. Masses are in units of M$_{\odot}$. For comparison, we also show the parameters for two previously reported events, GW170729~\citep{2019PhRvD.100j4015C} and GW190521~\citep{PhysRevLett.125.101102,2021PhRvX..11b1053A}, which were also detected by template-based searches, but were reported by cWB with the highest significance.}
\label{tab:PE}
\end{table*}
}

\section{Astrophysical Interpretation}

To gain a better understanding of where one can expect cWB to detect events that template-based searches miss, we can look at the binary mergers that have been previously detected by cWB with higher significance. There are two such events, both black hole mergers: GW170729 \citep{2019PhRvD.100j4015C} and GW190521 \citep{PhysRevLett.125.101102}. We show the reconstructed binary properties of these two events in Table \ref{tab:PE}. We see that GW170729 and GW190521 are among the heaviest black hole mergers detected by LIGO and Virgo. This is not surprising as cWB performs the best compared to the template-based searches at high masses \citep{2022A&A...659A..84A}. The GW190521 has a different astrophysical origin compared to the stellar mass BBH population detected by LIGO and Virgo \citep{LIGOScientific:2020ufj}. Both signals may be affected by precession or residual eccentricity diminishing their detection by the template searches. 

Detection of the 3 new cWB-only events may hint at unusual binary mergers. But our PE studies do not support this deduction. All three events are in the bulk of the stellar mass distribution, except for 190711 which has an unusually high mass ratio. 
Using the stellar-mass BBH simulations in GWTC-3 (refer \ref{subsec:sim}) which follow the astrophysical BBH population~\citep{2023PhRvX..13a1048A}, we compare the detection efficiency of the upgraded cWB pipeline with the CBC pipelines PyCBC~\citep{DalCanton:2020vpm}, GstLAL~\citep{PhysRevD.108.043004}, and MBTA~\citep{Aubin:2020goo}) used in O3. We find that the CBC searches perform best for the low chirp mass events, while cWB surpasses the CBC detection efficiency for ${\cal M}_{\rm c} > 30$\,M$_\odot$. From this simulation study, we estimate the fraction of the stellar-mass black hole mergers that are detected by cWB with high significance (IFAR\,$\geq 4.5\,$yr) while they are not detected or detected with low significance (IFAR\,$<1$\,yr) by any template-based search used in the O3 run. We find that cWB detects $\sim7.4\%$ of such BBH mergers that are missed by the other CBC searches. From the total number of 62 events detected by all pipelines in O3 with IFAR\,$>1$\,yr \citep{2023PhRvX..13d1039A}, we estimate the expected number of cWB-only events to be $4.8\pm{2.1}$. This is consistent with the cWB detections, indicating that the 3 cWB-only events likely belong to the observed stellar-mass BBH population.

Additionally, we note that the events 190711 and 200318 have been independently reported by non-LVK GW searches for BBH mergers. The event 190711 was found by the IAS search \citep{PhysRevD.106.043009, Wadekar:2023gea} with a comparable IFAR of $38.7$~yr. The event 200318 was found with a lower significance as compared to the cWB search, with IFAR $= 0.66$ yr in the IAS catalog \citep{Wadekar:2023gea} and IFAR $= 0.5$ yr in the 4-OGC catalog \citep{Nitz:2021zwj}. The event 190607 was not found as significant in any GW catalog.

\section{Conclusion}
In this work, we presented the updated cWB search used for reanalysis of the O3 data. The updated cWB search outperforms the previous cWB version by being $\sim39\%$ more sensitive to BBH mergers, and detects 36 events with IFAR $>1$ yr in the O3 run. 

The new search identified 3 new GW events that were not previously reported in the GWTC-3 catalog \citep{2023PhRvX..13d1039A}. These events, with a combined significance of 3.6 $\sigma$, represent a statistical excess of events detected only by the burst search. Using a simulation study we showed that cWB has a higher detection rate for signals with the chirp mass ${\cal M}_{\rm c} > 30$\,M$_\odot$, and detects $\sim7.4\%$ stellar-mass BBH mergers that can be missed by the template-based searches.

The 3 new candidate events are consistent with the expected number of cWB-only BBH detections in O3 ($4.8 \pm 2.1$), and belong to the stellar-mass binary population with the total masses in the $70-100$\,M$_\odot$ range. This highlights the cWB search's importance in exploring the high-mass end of the black hole mass spectrum. Notably, the highest significance event 190711 with $m_1=69.9^{+35.6}_{-18.3}$\,M$_\odot$ suggests a black hole in the pair-instability mass gap, and shows highly asymmetric masses ($q={0.29}^{+0.17}_{-0.13}$) indicative of a potential dynamical or AGN-assisted formation scenario.

With the ongoing fourth observing run of LIGO/Virgo/KAGRA, we expect to detect more cWB-only events hinting at an emerging population of unique BBH systems that might be missed by template-based searches.

\begin{acknowledgments}

This research has made use of data, software, and/or web tools obtained from the Gravitational Wave Open Science Center, a service of LIGO Laboratory, the LIGO Scientific Collaboration, and the Virgo Collaboration. We gratefully acknowledge the support of LIGO and Virgo for the provision of computational resources, especially LIGO Laboratory, which is supported by the NSF Grant No. PHY 0757058 and PHY 0823459. This work was supported by the NSF Grants No. PHY 2110060, PHY 2409372, and PHY 2309024. G.V. acknowledges the support of the NSF under grant PHY-2207728. M.S. acknowledges Polish National Science Centre Grant No. UMO-2023/49/B/ST9/02777 and the Polish National Agency for Academic Exchange within Polish Returns Programme Grant No. BPN/PPO/2023/1/00019. We acknowledge the valuable feedback by Archana Pai and Bence Kocsis. We acknowledge the use of open source Python packages including \textsc{NumPy}~\cite{numpy}, \textsc{Pandas}~\cite{pandas}, \textsc{Matplotlib}~\cite{matplotlib}, and \textsc{scikit-learn}~\cite{scikit-learn}.
\end{acknowledgments}

\appendix

\begin{table*}[bht]
\centering
    \setlength{\tabcolsep}{4pt}

  \begin{tabular}{lcccc}
        \hline
        \hline
         Event & \multicolumn{1}{c}{Standard cWB} & \multicolumn{1}{c}{ Updated cWB} & SNR & $P_{\mathrm{astro}}$ \\
         & \multicolumn{1}{c}{FAR [yr$^{-1}$]} & \multicolumn{1}{c}{FAR [yr$^{-1}$]} & & \\[0.5ex]
        \hline
        \hline 
        GW190408\_181802 & $< 9.5 \times 10^{-4}$ & $< 2.5 \times 10^{-3}$ & $16.1$ & $0.9997$ \\
        GW190412 & $< 9.5 \times 10^{-4}$ & $< 2.5 \times 10^{-3}$ & $20.9$ & $0.9999$\\
        GW190413\_134308 & \multicolumn{1}{c}{$...$} & $\quad\,1.7 \times 10^{-1}$ & $9.6$ & $0.9130$\\
        GW190421\_213856 & $\quad\, 3.0 \times 10^{-1}$ & $< 2.2 \times 10^{-3}$ & $11.3$ & $0.9991$\\
        GW190503\_185404 & $\quad\,1.8 \times 10^{-3}$ & $\quad\,2.2 \times 10^{-3}$ & $12.2$ & $0.9965$\\
        GW190512\_180714 & $\quad\,3.0 \times 10^{-1}$ & $< 2.2 \times 10^{-3}$ & $13.0$ & $1.000$\\
        GW190513\_205428 & \multicolumn{1}{c}{$...$} & $< 2.2 \times 10^{-3}$ & $13.2$ & $1.000$\\
        GW190517\_055101 & $\quad\,6.5 \times 10^{-3}$ & $\quad\,3.7 \times 10^{-2}$ & $11.2$ & $0.9922$\\
        GW190519\_153544 & $\quad\,3.1 \times 10^{-4}$ & $< 2.2 \times 10^{-3}$ & $14.8$ & $0.9999$\\
        GW190521 & $\quad\,2.0 \times 10^{-4}$ & $< 2.2 \times 10^{-3}$ & $14.7$ & $1.000$\\
        GW190521\_074359 & $< 1.0 \times 10^{-4}$ & $< 2.2 \times 10^{-3}$ & $24.9$ & $0.9997$\\
        GW190527\_092055 & \multicolumn{1}{c}{$...$} & $\quad\,5.1 \times 10^{-1}$ & $9.0$ & $0.7630$\\
        GW190602\_175927 & $\quad\,1.5 \times 10^{-2}$ & $< 2.3 \times 10^{-3}$ & $12.2$ & $0.9997$\\
        GW190701\_203306 & $\quad\,5.5 \times 10^{-1}$ & $\quad\,9.8 \times 10^{-3}$ & $10.1$ & $0.9891$\\
        GW190706\_222641 & $< 1.0 \times 10^{-3}$ & $< 2.5 \times 10^{-3}$ & $13.0$ & $0.9998$\\
        GW190707\_093326 & \multicolumn{1}{c}{$...$} & $\quad\,4.9 \times 10^{-3}$ & $11.9$ & $0.9949$\\
        GW190727\_060333 & $\quad\,8.8 \times 10^{-2}$ & $< 2.5 \times 10^{-3}$ & $11.6$ & $0.9998$\\
        GW190728\_064510 & \multicolumn{1}{c}{$...$} & $< 2.5 \times 10^{-3}$ & $12.9$ & $0.9998$\\
        GW190803\_022701 & \multicolumn{1}{c}{$...$} & $\quad\,4.6 \times 10^{-1}$ & $8.8$ & $0.9195$\\
        GW190828\_063405 & $< 9.6 \times 10^{-4}$ & $< 2.2 \times 10^{-3}$ & $17.3$ & $0.9998$\\
        GW190915\_235702 & $< 1.0 \times 10^{-3}$ & $\quad\,2.2 \times 10^{-3}$ & $14.7$ & $0.9973$\\
        GW190929\_012149 & \multicolumn{1}{c}{$...$} & $\quad\,2.8 \times 10^{-2}$ & $11.0$ & $0.9862$\\
        GW191109\_010717 & $< 1.1 \times 10^{-3}$ & $< 2.3 \times 10^{-3}$ & $16.9$ & $0.9998$\\
        GW191204\_171525 & $< 8.7 \times 10^{-4}$ & $< 2.4 \times 10^{-3}$ & $18.0$ & $0.9998$\\
        GW191215\_223052 & $\quad\,1.2 \times 10^{-1}$ & $< 2.4 \times 10^{-3}$ & $11.3$ & $0.9999$\\
        GW191222\_033537 & $< 8.9 \times 10^{-4}$ & $\quad\,2.4 \times 10^{-3}$ & $12.5$ & $0.9974$\\
        GW191230\_180458 & $\quad\,5.0 \times 10^{-2}$ & $\quad\,2.4 \times 10^{-3}$ & $11.0$ & $0.9972$\\
        GW200128\_022011 & $\quad\,1.3 \times 10^{+0}$ & $\quad\,2.8 \times 10^{-1}$ & $10.9$ & $0.9031$\\
        GW200208\_130117 & \multicolumn{1}{c}{$...$} & $\quad\,9.2 \times 10^{-3}$ & $10.0$ & $0.9964$\\
        GW200209\_085452 & \multicolumn{1}{c}{$...$} & $\quad\,4.6 \times 10^{-3}$ & $10.5$ & $0.9969$\\
        GW200219\_094415 & $\quad\,7.7 \times 10^{-1}$ & \multicolumn{1}{c}{$...$} & $..$ & $..$\\
        GW200224\_222234 & $< 8.8 \times 10^{-4}$ & $< 2.5 \times 10^{-3}$ & $18.4$ & $0.9998$\\
        GW200225\_060421 & $< 8.8 \times 10^{-4}$ & $< 2.5 \times 10^{-3}$ & $13.6$ & $0.9999$\\
        GW200311\_115853 & $< 8.2 \times 10^{-4}$ & $< 2.3 \times 10^{-3}$ & $16.4$ & $0.9998$\\
        
        \hline
    \end{tabular}
    \caption{O3 detections by the updated cWB search are compared with the standard cWB search results. We report all the cWB detections with FAR $\leq1\, \mathrm{yr}^{-1}$ and present in the GW transient catalog~\cite{2023PhRvX..13d1039A}. The estimated significance is limited by the accumulated background data and is indicated with a '$<$' entry. We estimate the $P_{\mathrm{astro}}$ and report it along with the SNR reconstructed by the updated cWB pipeline for the GW events.}
\label{tab:O3events}
\end{table*}

\newpage

\bibliography{Refs}
\bibliographystyle{aasjournal}

\end{document}